%% file: main.tex
\PassOptionsToPackage{table,xcdraw}{xcolor}
\documentclass[sigconf, screen]{acmart}

\AtBeginDocument{%
  }

\setcopyright{none} 
\copyrightyear{2025}
\acmYear{2025}
\acmDOI{XXXXXXX.XXXXXXX}

\acmISBN{978-X-XXXX-XXXX-X/XX/XX}

\usepackage{subfig}
\usepackage{tikz}
\usepackage{enumitem}
\usepackage{pifont}
\newcommand{\cmark}{\ding{51}}%
\newcommand{\xmark}{\ding{55}}%
\usepackage{tabularx}
\usepackage{multirow}
\usepackage{threeparttable}
\usepackage{tcolorbox}

\usepackage{makecell}
\usepackage{enumitem}
\usepackage{booktabs}
\usepackage{colortbl}
\usepackage{hhline}
\usepackage{subfig}
\usepackage{graphicx,scalerel}
\usepackage{textgreek}
\usepackage{caption}
\usepackage{etoolbox}

\AtBeginEnvironment{equation}{%
  \setlength{\abovedisplayskip}{2pt}%
  \setlength{\belowdisplayskip}{2pt}%
  \setlength{\abovedisplayshortskip}{2pt}%
  \setlength{\belowdisplayshortskip}{2pt}%
}
\AtBeginEnvironment{align}{%
  \setlength{\abovedisplayskip}{2pt}%
  \setlength{\belowdisplayskip}{2pt}%
  \setlength{\abovedisplayshortskip}{2pt}%
  \setlength{\belowdisplayshortskip}{2pt}%
}

\definecolor{mypurple}{HTML}{D5C9DF}
\definecolor{dmgreen}{HTML}{169136}

\newcommand{\AT}[1]{\textcolor{black}{#1}}

\title{Leveraging Core and Uncore Frequency Scaling for Power-Efficient Serverless Workflows
}

\author[]{Achilleas Tzenetopoulos, Dimosthenis Masouros, Sotirios Xydis, Dimitrios Soudris}
\affiliation{\institution{National Technical University of Athens}\country{}\city{}}

\begin{document}

\begin{abstract}

Serverless workflows have emerged in Function-as-a-Service (FaaS) platforms to represent the operational structure of traditional applications.
With latency propagation effects becoming increasingly prominent, step-wise resource tuning is required to address Service-Level-Objectives (SLOs).
Modern processors' allowance for fine-grained Dynamic Voltage and Frequency Scaling (DVFS), coupled with serverless workflows' intermittent nature, presents a unique opportunity to reduce power while meeting SLOs.
We introduce $\Omega$kypous, an SLO-driven DVFS framework for serverless workflows. $\Omega$kypous employs a grey-box model that predicts functions' execution latency and power under different Core and Uncore frequency combinations. Based on these predictions and the timing slacks between workflow functions, $\Omega$kypous uses a closed-loop control mechanism to dynamically adjust Core and Uncore frequencies, thus minimizing power consumption without compromising predefined end-to-end latency constraints.
Our evaluation on real-world traces from Azure, against state-of-the-art power management frameworks, demonstrates an average power consumption reduction of 16\%, while consistently maintaining low SLO violation rates (1.8\%), when operating under power caps.

\end{abstract}

  
\maketitle


\input{sections/00_introduction}
\input{sections/01_related}
\input{sections/02_motivation}
\input{sections/03_problem}
\input{sections/04a_okypous_design}
\input{sections/04b_okypous_modeling}
\input{sections/04c_okypous_resolver}
\input{sections/04d_okypous_implementation}
\input{sections/05a_evaluation_setup}
\input{sections/05b_evaluation_results}
\input{sections/06_conclusion}



\bibliographystyle{ACM-Reference-Format}
\bibliography{refs}

\end{document}

%% file: sections/00_introduction.tex
\section{Introduction}
\label{sec:intro}

Serverless workflows extend the Function-as-a-Service (FaaS) paradigm by allowing serverless functions to execute in a structured manner based on predefined execution dependencies~\cite{zhou2022aquatope,stojkovic2024ecofaas,burckhardt2022netherite}. 
Unlike independent function invocations, workflows provide mechanisms for sequencing, parallel execution, and event-driven coordination.
Major cloud providers offer workflow orchestration services--such as Azure Durable Functions~\cite{azure-durable}, AWS Step Functions~\cite{awsstep}, and Google Workflows~\cite{gcpwf}--which support large-scale applications, processing billions of workflows per day~\cite{burckhardt2022netherite,mahgoub2022wisefuse}.

Serverless workflow platforms allow developers to define applications as sequences of loosely coupled functions, each representing a distinct processing stage. These workflows are described using declarative formats (e.g., YAML, JSON) or imperative code (e.g., Python, JavaScript), specifying execution order, control logic, and event-driven triggers. 
While current platforms do not expose explicit Service Level Objectives (SLOs), recent work has proposed SLO-driven models to enable latency-aware control and resource adaptation~\cite{mahgoub2021sonic,zhou2022aquatope,moghimi2023parrotfish,stojkovic2024ecofaas}. These proposals introduce the idea of integrating latency or tail-latency targets into workflow specifications, thereby allowing systems to better align resource decisions with application-level goals.

The structured and decoupled nature of serverless workflows enables cloud providers to dynamically adjust resource allocation at function level.
Since functions differ in resource utilization profiles (e.g., CPU-, memory-bound), they can be dynamically "right-sized" to match their non-functional requirements.
While prior work has explored this flexibility optimizing SLOs~\cite{mahgoub2021sonic,zhou2022aquatope,moghimi2023parrotfish,seqclock, wen2022stepconf,wu2024faasbatch}, power management has received significantly less attention~\cite{stojkovic2024ecofaas,tzenetopoulos2023dvfaas}.
Yet, power consumption is an increasingly critical concern for data center operators~\cite{lo2014towards, haque2017exploiting, li2022towards}, as it impacts both operational costs and compliance with power budget contracts~\cite{zhang2021flex}.
Notably, cloud providers are beginning to incorporate power-aware policies: Google, for instance, factors CPU speed into its billing model~\cite{google-freq}, while IBM has recently released Kepler, a cloud-native power monitoring framework~\cite{amaral2023kepler}, highlighting the growing demand for more granular power management solutions.

\noindent\textbf{\underline{Power management strategies:}} Cloud infrastructures typically manage power consumption using  Dynamic Voltage and Frequency Scaling (DVFS) techniques ~\cite{retail, gemini2020zhou, zhang2023first, liu2024improving, stojkovic2024ecofaas, tzenetopoulos2023dvfaas}, the utilization of Asymmetric Multicore Processors (AMPs) ~\cite{haque2017exploiting} or the integration of sleep state management ~\cite{chou2019mudpm, sharafzadeh2019yawn, agilewatts}.
Among these methods, DVFS has become particularly prominent, initially targeting microservice deployments~\cite{retail,gemini2020zhou,kasture2015rubik} and, more recently, serverless workflows~\cite{tzenetopoulos2023dvfaas,stojkovic2024ecofaas}.
However, these approaches exclusively optimize the Core frequency scaling (CFS), ignoring the potential power savings offered by adjusting the Uncore frequency (UFS).
As modern processor architectures (e.g., post Intel\textsuperscript{\textregistered} Haswell) support independent scaling of Core and Uncore frequencies, there is a largely unexplored opportunity for jointly optimizing CFS and UFS to further improve power efficiency.
Our study (\S\ref{sec:challenges}) shows that UFS can yield power reductions of up to 31\%, providing additional headroom for power efficiency optimizations.

\input{tables/related}

\noindent \textbf{\underline{Harnessing the ``\textit{power}'' of serverless functions:}} Realizing these power-efficiency benefits in practice requires adaptive execution aligned with the workflow’s SLO.
Given the complete visibility of the workflow's structure, cloud providers can leverage \textit{timing slacks}~\cite{bhasi2021kraken, zhang2025slack}, i.e., deviations between allocated and actual execution time of preceding functions in the workflow, to optimize resource usage dynamically.
Since slacks emerge unpredictably and propagate across function stages, \textit{runtime control} becomes essential.
The system must be able to adapt to the evolving latency budget throughout the workflow by adjusting CFS and UFS on a per-invocation basis.
While monitoring timing deviations is necessary, it is not sufficient, since reactive adjustments alone cannot prevent SLO violations.
To proactively avoid them, the system requires accurate latency prediction ahead of execution.
This necessitates fine-grained performance modeling at the function level, capable of capturing execution behaviors under diverse CFS\&UFS settings~\cite{sahraei2023xfaas,shahrad2019architectural,romero2021faa}.
Moreover, inference must be lightweight, as predictions occur in the critical path of function scheduling.
While prior works have explored various modeling techniques for serverless and microservices, these approaches either depend on costly per-function profiling~\cite{retail,stojkovic2024ecofaas}, assume fixed latency-frequency relationships~\cite{gemini2020zhou}, or require extensive iterations to identify efficient configurations~\cite{zhou2022aquatope}.
Notably, according to Azure's insights~\cite{zhang2021faster}, invocation counts in serverless workflows are often insufficient to train ML-based latency estimation models, highlighting the need for lightweight, data-efficient alternatives.

\noindent\textbf{\underline{Our work.}} This paper presents $\Omega$kypous, a power-aware, SLO-driven control framework for serverless workflows. $\Omega$kypous dynamically adjusts Core and Uncore frequencies at the granularity of individual function invocations, leveraging timing slacks that naturally emerge during workflow execution. Unlike prior systems that rely on per-function profiling, $\Omega$kypous uses a lightweight, global grey-box model to predict latency across unseen functions and configurations. This model, paired with a slack-aware controller, enables \textit{``just-right''} frequency scaling, improving power efficiency without compromising performance.
Our main contributions are:
\begin{enumerate}[leftmargin=10pt,nosep,label=\roman*.]
    \item We offer a detailed discussion on serverless DVFS strategies and opportunities, showing the effects of CFS\&UFS on the latency and power consumption of serverless functions (\S\ref{sec:challenges}).
    \item We introduce $\Omega$kypous (\S\ref{sec:design}), an adaptive frequency controller for SLO-aware serverless workflows that exploits inter-function slack propagation to minimize power consumption.
    To the best of our knowledge, $\Omega$kypous is the first serverless framework that jointly exploits CFS\&UFS for power-efficient execution.
    \item We propose a global grey-box model (\S\ref{sec:grey-box}) for estimating function latency that combines first-principles knowledge with data-driven regression components, along with an explainability study of its features. 
    We show that, unlike previous approaches, which require per-frequency latency modeling~\cite{retail}, the latency of unseen functions can be estimated with minimal samples ($\approx$ 5) and an overall Mean Absolute Percentage Error (MAPE) of $\approx 4\%$.
    \item We provide a qualitative (\S\ref{sec:related}) and quantitative (\S\ref{sec:eval}) evaluation of $\Omega$kypous against sota DVFS frameworks, showing its benefits in both power savings and SLO compliance.
\end{enumerate}

\noindent Through extensive evaluation, we
demonstrate that $\Omega$kypous outperforms Linux's built-in governors and SotA implementations (\textit{DVFaaS}~\cite{tzenetopoulos2023dvfaas}, \textit{EcoFaaS}~\cite{stojkovic2024ecofaas},
\textit{Gemini}~\cite{gemini2020zhou} and \textit{ReTail}~\cite{retail}) in terms of SLO violations and power efficiency.
Notably, $\Omega$kypous achieves these improvements using a single global model, whereas prior approaches are evaluated with oracle latency predictors to isolate their control logic from modeling accuracy.
By conducting experiments on real serverless invocation traces from Azure Functions~\cite{zhang2021faster}, we showcase that $\Omega$kypous achieves to reduce the SLO violation rate to 1.8\% and improves power efficiency by up to 22\% over the best-performing baselines~\cite{tzenetopoulos2023dvfaas,stojkovic2024ecofaas}.

%% file: tables/related.tex
\begin{table*}[t]
\centering
\caption{
Qualitative comparison between state-of-the-art and $\Omega$kypous}
\scriptsize
\begin{threeparttable}
\begin{tabularx}{0.8\textwidth}{ccccccc|c}
\hline
\centering
\textbf{}&\textbf{Gemini}\cite{gemini2020zhou}&\textbf{ReTail}\cite{retail}&\textbf{SeqClock}\cite{seqclock}&\textbf{Aquatope}\cite{zhou2022aquatope}&\textbf{DVFaaS}\cite{tzenetopoulos2023dvfaas}&\textbf{EcoFaaS}\cite{stojkovic2024ecofaas}&\textbf{$\Omega$kypous}\\
\hline
\rowcolor{mypurple}\textbf{Modeling}& Per App& Per App $\times$ Freq. lvl & -&Per Workflow $\times$ Input&-&Per App $\times$ Freq. lvl& Global\\
\textbf{Objective (s.t. SLO)}&Power Min.&Power Min.& Power Min. & CPU util. Min. &Cost Min.& Power Min.&Power Min.\\
\rowcolor{mypurple}\textbf{Technique}&RAPL&CFS&CFS&Core Scaling&Core Scaling&CFS&CFS \& UFS\\
\textbf{Scope}&Service&Service&Sequence& Workflow& Sequence&Sequence&Workflow\\
\rowcolor{mypurple}\textbf{Propagation-aware}&\xmark&\xmark &\cmark & \xmark&\cmark&\cmark&\cmark\\
\hline
\end{tabularx}
\end{threeparttable}
\label{tab:related}
\end{table*}

%% file: sections/01_related.tex
\section{Related Work}
\label{sec:related}

\noindent$\blacktriangleright$ \textbf{SLO-aware strategies}: 
Several prior works optimize resource allocation for latency-critical applications under SLO constraints~\cite{chen2019parties,patel2020clite,nishtala2020twig,zhang2021sinan}, but they do not focus on power management.
Recent research on serverless workflows targets cost minimization ~\cite{moghimi2023parrotfish,wen2022stepconf,wu2024faasbatch, zhou2022aquatope} or end-to-end latency regulation ~\cite{seqclock,zhao2021understanding,da2019using, bhasi2021kraken,mahgoub2022orion,sadeghian2023unfaasener}.
SeqClock\cite{seqclock} and Aquatope\cite{zhou2022aquatope} are closely related to our work. 
They employ CPU-quota scaling to meet SLO requirements in serverless workflows but do not aim to minimize power.
SeqClock follows a function-agnostic, step-wise, closed-loop approach to regulate end-to-end execution latency, while Aquatope \cite{zhou2022aquatope} employs Bayesian Optimization to minimize CPU time usage while meeting SLO constraints.
However, as we later show in \S\ref{sec:challenges}, DVFS strategies appear to be a more suitable configuration knob for power management compared to CPU quotas.

\noindent$\blacktriangleright$ \textbf{DVFS-enabled power management}: 
Several studies have aimed to optimize the power consumption of latency-critical applications using DVFS~\cite{retail,lo2014towards,kasture2015rubik,gemini2020zhou,tzenetopoulos2023dvfaas, stojkovic2024ecofaas, zhang2023first}.
Among these works, Gemini~\cite{gemini2020zhou} and ReTail~\cite{retail} are most closely related to ours. Both propose CFS-driven strategies at the microservice level, but neglect slack propagation effects across the entire service.
Gemini \cite{gemini2020zhou} relies on application-specific neural networks for request latency prediction, which can introduce huge overheads to the milliseconds requirements of serverless. 
In addition, it operates under the assumption of linear proportionality between frequency and latency.
On the other hand, Retail~\cite{retail} relies on linear regression techniques, which, although simpler and more generalizable, require expensive modeling (i.e., extensive profiling to model every possible frequency configuration).
Despite their respective modeling limitations (\S\ref{subsec:latency_and_power_prediction}), their focus remains on CFS-driven mechanisms, leaving the impact of UFS on power management relatively unexplored. 
Recently, DVFaaS\cite{tzenetopoulos2023dvfaas}, and EcoFaaS \cite{stojkovic2024ecofaas} have been introduced that capitalize on the intermittent nature of serverless workflows to address their end-to-end latency requirements while minimizing power consumption.
However, they still rely on CFS-driven power management and inadequately manage noise, either due to the absence of latency modeling \cite{tzenetopoulos2023dvfaas} or coarse-grained, pool-based DVFS \cite{stojkovic2024ecofaas}.

\noindent$\blacktriangleright$ \textbf{$\Omega$kypous value \& differentiators}: 
$\Omega$kypous introduces a DVFS approach for serverless workflows using a single, lightweight global grey-box model to predict function execution time, avoiding costly per-function profiling~\cite{retail} or fixed proportional frequency-latency mappings~\cite{gemini2020zhou}. 
Unlike prior approaches that focus solely on CFS~\cite{gemini2020zhou,retail,tzenetopoulos2023dvfaas,stojkovic2024ecofaas}, $\Omega$kypous also accounts for the impact of UFS, unlocking additional power-saving headroom that would otherwise remain unexploited.
To handle dynamic execution variability, it incorporates an adaptive controller and leverages the inherent modularity of serverless workflows to coordinate slack propagation across function invocations. Table~\ref{tab:related} summarizes the key differences between $\Omega$kypous and state-of-the-art DVFS frameworks.

%% file: sections/02_motivation.tex
\section{Power Efficiency in Serverless Workflows: Opportunities and Challenges}
\label{sec:challenges}
In this section, we examine the suitability of leveraging Core (CFS) and Uncore (UFS) frequency scaling to enhance the power efficiency of serverless functions. 
Specifically, we analyze the potential benefits of coordinated CFS\&UFS configurations compared to \textit{i)} CFS-only approaches, \textit{ii)} CPU-quota scaling strategies proposed in prior work and \textit{iii)} existing Linux frequency governors. 
Moreover, we investigate \textit{timing slack} opportunities in serverless environments and the challenges associated with accurately modeling function performance under varying CFS–UFS configurations.
We conduct this analysis on a real system using various functions from the SeBs~\cite{sebs} benchmark suite, deployed on OpenWhisk\footnote{The detailed experimental setup can be found in \S \ref{sec:experimental-setup}.}.

\begin{figure}[t]
    \centering
    \includegraphics[width=\columnwidth]{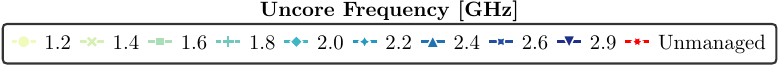}
\\
\vspace{2pt}
\includegraphics[width=\columnwidth]{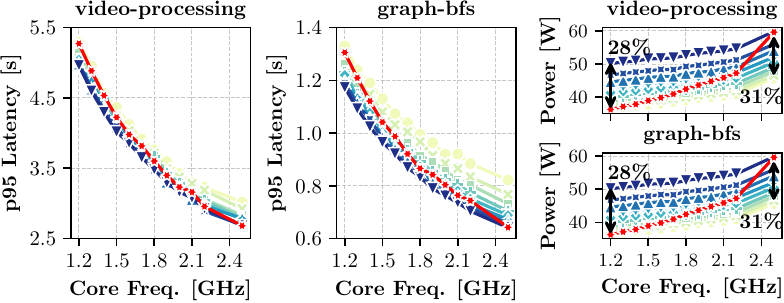}
    \caption{Impact of Core and Uncore frequency scaling on tail latency and power consumption.}
\label{fig:cfs-ufs-latency-power-analysis}
\end{figure}

\noindent$\blacktriangleright$ \textbf{Performance/Power benefits of CFS\&UFS:}
Figure~\ref{fig:cfs-ufs-latency-power-analysis} shows the performance and power consumption under different Core (x-axis) and Uncore (hued colors) configurations. 
CFS plays a dominant role in reducing tail latency, with higher frequencies consistently leading to lower execution times.
However, UFS also contributes to performance tuning, providing up to $\approx$20–30\% latency headroom for optimization, depending on the workload.
Notably, the impact of UFS is more pronounced in the \texttt{\small graph-bfs} function, which inherently involves more memory-intensive operations.
Finally, unmanaged UFS (red line) deviates from the controlled settings, indicating an adaptive DVFS approach, which seeks to balance performance and power dynamically.
Regarding power consumption, power increases monotonically with core frequency, with a steeper rise beyond 2.2 GHz.
Since latency does not exhibit a proportional decrease, this trend suggests that higher frequencies yield diminishing returns in performance improvements relative to power cost.
Overall, power consumption closely tracks the CFS-UFS settings, indicating that power behavior is more strongly influenced by the micro-architecture than by the workload characteristics of the examined functions.
Notably, scaling Core frequency alone from 1.2GHz to 2.5GHz offers power savings headroom ranging from 15\% (Uncore=1.2GHz) to 17\% (Uncore=2.9GHz), whereas scaling Uncore frequency results in more substantial power reductions of 28\% (Core=1.2GHz) to 31\% (Core=2.5GHz).

\vspace{-5pt}
\begin{tcolorbox}[colback=gray!15, colframe=black!100, boxrule=0.1mm, arc=1mm, left=2pt, right=2pt, top=0pt, bottom=0pt]
\textit{$\star$\textbf{Takeaways:} 1) Core frequency is the main knob for tuning latency, while Uncore frequency is more effective for power savings. 
2) Joint CFS \& UFS optimization provides additional headroom for optimization.
3) Power draw is governed more by CFS-UFS settings than by workload variability in serverless environments.}
\end{tcolorbox}

\begin{figure}[t]
	\centering
    \centering
    \includegraphics[width=.85\columnwidth]{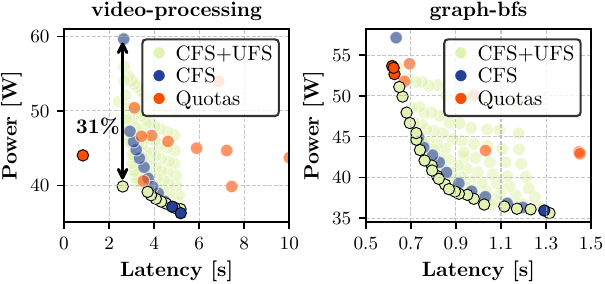}
     \caption{Pareto analysis of latency and power consumption trade-offs, between \textit{i)} CFS\&UFS (beige), \textit{ii)} CFS only (blue) and \textit{iii)} CPU Quotas (orange).}
     \label{fig:pareto-analysis}
\end{figure}
\noindent$\blacktriangleright$ \textbf{Performance/Power Pareto analysis:} 
To further illustrate the benefits of a joint CFS\&UFS configuration, we examine the Pareto frontier of latency and power trade-offs. 
In addition to frequency scaling, we explore CPU quotas as an alternative control mechanism, which have been leveraged in prior works to regulate performance of serverless workflows~\cite{seqclock,zhou2022aquatope}.
CPU quotas constrain the processing time allocated to a function, directly impacting performance while indirectly modulating power consumption through CPU utilization control.
We consider CPU quotas ranging from 0.25 to the maximum number of cores, in increments of 0.25.
Figure~\ref{fig:pareto-analysis} presents the design space, where the opaque points denote Pareto-optimal configurations, i.e., no other configurations achieve both lower power and better performance simultaneously.
We observe that joint CFS\&UFS tuning not only expands the Pareto frontier but also dominates the majority of configurations.
Compared to a CFS-only approach, CFS+UFS provides significant power savings, particularly at lower latencies, reducing power consumption by up to 33\% in an iso-latency scenario for the \texttt{\small video-processing} benchmark.
Notably, since CFS\&UFS forms a superset of CFS, any CFS-dominated points (primarily at higher latencies) are included within the CFS\&UFS space.
Regarding CPU quotas, we observe that they can offer lower-latency Pareto-optimal configurations, but at higher power costs, due to Linux governor's dynamic frequency scaling.
Notably, unlike CFS\&UFS, which forms a structured Pareto frontier, CPU quotas distribution is highly scattered, marking them as a less predictable knob for power-efficient performance tuning.

\vspace{-5pt}
\begin{tcolorbox}[colback=gray!15, colframe=black!100, boxrule=0.1mm, arc=1mm, left=2pt, right=2pt, top=0pt, bottom=0pt]
\textit{$\star$\textbf{Takeaways:} 1) Joint CFS\&UFS configuration expands the Pareto frontier, offering more power-efficient configurations for iso-latency scenarios compared to CFS. 2) CPU quotas exhibit scattered behavior, making them a less predictable knob for power-efficient performance tuning.}
\end{tcolorbox}

\noindent$\blacktriangleright$ \textbf{Viability of existing Linux governors:}
Modern processors support DVFS via CPU governors, software modules in the Linux kernel, to control DVFS based on system demand.
The four common CPU governors are i) \textit{powersave}, ii) \textit{performance}, iii) \textit{ondemand}, and iv) \textit{conservative}. 
To assess the impact of each governor on performance and power consumption, we sequentially invoke our examined functions four times, each time using one governor exclusively.
Our findings indicate that \textit{powersave} and \textit{performance} governors anchor the CPU frequency at its minimum and maximum limits, respectively. 
In contrast, \textit{ondemand} and \textit{conservative} governors adjust the CPU frequency based on CPU utilization, with \textit{conservative} responding more gradually to changes, promoting reduced power consumption (37, an 42\% respectively compared to \textit{performance}). 
However, these governors operate without visibility into application-level SLOs, and thus cannot adapt frequency based on latency targets. As a result, Linux DVFS governors are fundamentally SLO-agnostic and may miss opportunities for power-efficient execution under SLO constraints.
\vspace{-5pt}
\begin{tcolorbox}[colback=gray!15, colframe=black!100, boxrule=0.1mm, arc=1mm, left=2pt, right=2pt, top=0pt, bottom=0pt]
\textit{$\star$\textbf{Takeaways:} Linux DVFS governors provide power-saving benefits but lack SLO awareness, limiting their ability to optimize power efficiency without risking SLO violations.}
\end{tcolorbox}

\begin{figure}
    \centering
    \subfloat[Azure Functions \cite{zhang2021faster}.\label{fig:performance-variability-azure}]{
        \includegraphics[width=.7\columnwidth]{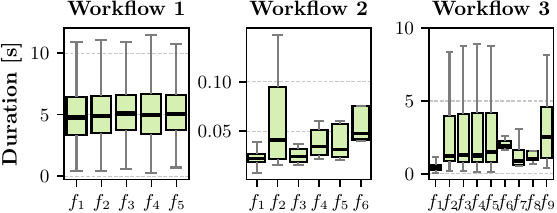}
    }
    \hspace*{\fill}
    \subfloat[AWS Lambda \cite{awslambda}.]{
        \includegraphics[width=0.25\columnwidth]{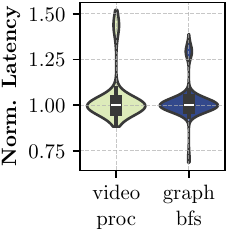}
        \label{fig:performance-variability-lambda}
    }
    \caption{Performance variability across function executions in serverless workflows.}
    \label{fig:function-slack}
\end{figure}

\noindent$\blacktriangleright$ \textbf{Timing slacks in serverless environments:} Serverless workflows exhibit timing variability due to a range of factors.
Execution time may fluctuate because of input-dependent behavior~\cite{retail}, shared infrastructure contention~\cite{zhao2021understanding} or cold-start effects~\cite{mohan2019agile}.
At the same time, functions might execute faster than expected~\cite{bhasi2021kraken} or even be conditionally entirely skipped~\cite{stojkovic2023specfaas}, depending on runtime logic.
Together, these factors create \textit{timing slacks}, i.e., deviations between the allocated and actual execution time of function stages.
Within the context of serverless workflows under end-to-end SLOs, such slacks present an opportunity for runtime optimization: by detecting them and dynamically performing CFS\&UFS, the system can reduce power consumption without violating end-to-end SLOs.

To examine the presence of timing slacks in practice, we analyze the performance variability of serverless functions across two real-world settings.
Figure~\ref{fig:performance-variability-azure} shows the distribution of function execution times across three workflows extracted from Azure's Functions trace~\cite{azure-durable}, with each box plot representing a distinct function and its latency distribution across multiple invocations.
We observe that function-level performance varies considerably both across and within workflows.
In Workflow 1, all functions exhibit relatively stable behavior, with execution times clustered around the median and ranging from approximately $1s$ to $10s$.
In contrast, Workflows 2 and 3 show more heterogeneous behavior. 
Some functions show tight latency distributions (e.g., variability within 2\% of the median), while others experience high dispersion, with median latencies up to $5\times$ faster than their respective tail latencies.
This effect is more pronounced in Workflow 3, where several functions have median execution times around $2s$ but occasionally spike to $9s$. 
To further validate the generality of this behavior, we conduct a complementary analysis on AWS Lambda~\cite{awslambda}.
Figure~\ref{fig:performance-variability-lambda} shows the normalized (to median) execution time distributions of \texttt{\small video-processing} and \texttt{\small graph-bfs} functions, measured across 480 invocations throughout the day.
While most invocations cluster tightly around the median—within 10\% variability—we observe occasional high-latency spikes in both functions, which are attributed to cold-start effects.
Additionally, for \texttt{\small graph-bfs}, we observe low-tail latencies reaching up to 25\% faster than the median, indicating that timing slacks arise in both directions.

\vspace{-5pt}
\begin{tcolorbox}[colback=gray!15, colframe=black!100, boxrule=0.1mm, arc=1mm, left=2pt, right=2pt, top=0pt, bottom=0pt]
\textit{$\star$\textbf{Takeaways:} Performance variability and conditional execution in serverless workflows introduce \textit{timing slacks}, which are propagated across function stages and can be exploited for power-efficient execution under SLO constraints.}
\end{tcolorbox}

\begin{figure}
    \centering
    \subfloat[Latency vs. Input]{
        \includegraphics[width=.62\columnwidth]{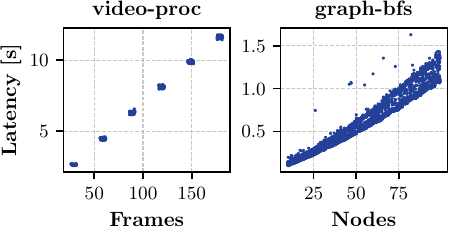}
        \label{fig:input-size-variability}
    }
    \subfloat[Invocation count \cite{zhang2021faster}.]{
        \includegraphics[width=0.37\columnwidth]{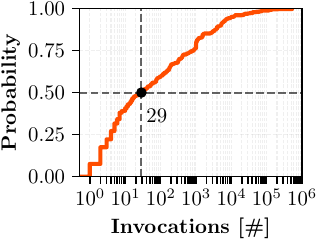}
        \label{fig:invocation-count}
    }
    \caption{Latency modeling challenges.}
    \label{fig:aaaaaaaa}
\end{figure}

\noindent$\blacktriangleright$ \textbf{Performance Modeling Challenges:} While timing slacks offer valuable opportunities for power optimization, realizing their benefits requires accurate prediction of function execution latency under varying CFS-UFS configurations. Such predictions are critical for runtime frequency scaling and latency budgeting, ensuring compliance with end-to-end SLOs.
However, performance modeling in serverless environments faces two challenges. 
First, functions are invoked with a wide range of input parameters, which can lead to significant and often nonlinear latency variation.
As shown in Figure~\ref{fig:input-size-variability} latency increases with input size (e.g., with the number of graph nodes in \texttt{graph-bfs} and the number of video frames in \texttt{video-proc}), highlighting the need for input-aware, generalizable models that can capture these patterns without per-function tuning.
Second, serverless environments often suffer from data sparsity, i.e., functions may be invoked only a few times, limiting the availability of training data for ML-based models.
Notably, insights from Azure's production trace (Figure~\ref{fig:invocation-count}) reveal that over half of all functions are invoked fewer than 30 times, highlighting the existence of ``few-shot'' scenarios.
While prior works have proposed ML-based predictors~\cite{retail,stojkovic2024ecofaas,zhou2022aquatope}, they train a separate model per function, which can fall short in small-data scenarios that are common in serverless environments.
Global models provide a more scalable alternative by enabling knowledge sharing across functions.
However, approaches based on complex neural networks often require large training datasets and incur high inference overhead, making them unsuitable for serverless systems, where predictions are performed in the critical path of function scheduling.

\vspace{-5pt}
\begin{tcolorbox}[colback=gray!15, colframe=black!100, boxrule=0.1mm, arc=1mm, left=2pt, right=2pt, top=0pt, bottom=0pt]
\textit{$\star$\textbf{Takeaways:} Serverless environments are characterized by input diversity and sparse invocation patterns, motivating the need for lightweight and input-aware global models that enable fast inference in the critical path of workflow execution.}
\end{tcolorbox}

%% file: sections/03_problem.tex
\section{$\Omega$kypous Overview}
\label{sec:formulation}

\noindent$\blacktriangleright$ \textbf{$\boldsymbol{\Omega}$kypous optimization goal:} 
Power consumption is a critical concern for modern cloud infrastructures, with power capping being a first-class strategy followed in high-end data centers \cite{lo2014towards,273871}.
$\Omega$kypous aims to minimize the power consumption of serverless workflows without violating their end-to-end SLO constraints.
While the global objective is to minimize power across the full workflow, solving this problem in practice requires full knowledge of execution paths and function runtimes in advance.
However, such an assumption does not hold in serverless environments, where control flow and execution times are dynamic and input-dependent.
As a result, it is impractical to determine an optimal power schedule across all function invocations prior to workflow execution (e.g., through MILP~\cite{stojkovic2024ecofaas}).
To address this, $\Omega$kypous adopts a \textit{performance-first, dynamic approach at function level: For each function comprising the workflow, it ensures that its latency budget is always met, and then opportunistically minimizes power consumption when timing slacks allow}.

\begin{figure}[t]
\centering
\includegraphics[width=.93\columnwidth,keepaspectratio=true]{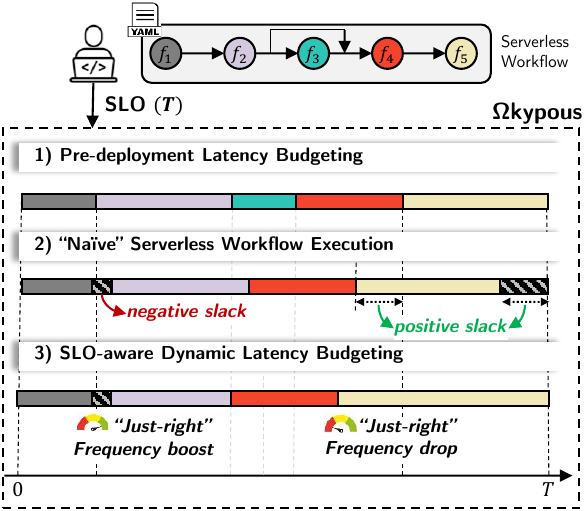}
\caption{$\boldsymbol{\Omega}$kypous optimization approach overview.}
\label{fig:motivational-exmample}
\end{figure}

\noindent$\blacktriangleright$ \noindent\textbf{Workflow Topology \& SLO acquisition:} 
As shown at the top of Figure ~\ref{fig:motivational-exmample}, and in line with production-level serverless workflow platforms ~\cite{awsstep,gcpwf}, $\Omega$kypous operates with full visibility into the application’s control flow.
The workflow topology is typically provided by developers through declarative specifications (e.g., YAML, JSON) or imperative APIs.
In addition to the workflow structure, and following common practices in prior works~\cite{stojkovic2024ecofaas,zhou2022aquatope}, developers also specify an end-to-end SLO constraint (e.g., p95th latency $\leq$ 1s).

\noindent$\blacktriangleright$ \textbf{$\boldsymbol{\Omega}$kypous optimization approach.}
Given complete visibility into the workflow structure and SLO targets, $\Omega$kypous capitalizes on runtime \textit{timing slacks}, to properly configure Core and Uncore frequency at function-level.
Timing slacks can be either \textit{i) negative} indicating delays caused by unpredictable runtime effects such as resource contention~\cite{chen2019parties,zhao2021understanding} or latency mispredictions~\cite{zhou2022aquatope,gemini2020zhou}, or \textit{ii) positive}, 
reflecting time gained due to early function completions, conditional control flow paths~\cite{stojkovic2023specfaas}, loose SLO constraints, or similar latency mispredictions.
Figure~\ref{fig:motivational-exmample} shows a high-level example of $\Omega$kypous' optimization approach for a serverless workflow comprising five functions and a conditional branch.
In this example, the user-specified SLO is defined based on the worst-case execution path, i.e., the cumulative latency of all functions along the critical path.
The process begins with an initial pre-deployment latency budgeting phase (\S\ref{subsubsec:predeployment-latency-budgeting}), where a nominal latency budget is statically assigned to each function.
However, during actual execution, naïve adherence to these initial budgets can lead to both negative slacks (e.g., if $f_1$ runs slower than expected) and positive slacks (e.g., if $f_3$ is conditionally skipped), resulting in variability in total execution time.
To adapt to such dynamic conditions, $\Omega$kypous monitors runtime timing slacks (\S\ref{subsubsec:dynamic-latency-budgeting}) and utilizes an adaptive controller (\S\ref{sec:controller}) to dynamically reassign latency budgets to subsequent functions.
Given a new latency budget, $\Omega$kypous employs a global performance model (\S\ref{sec:grey-box}) to predict the function’s expected latency under different CFS\&UFS configurations and 
selects the lowest-power CFS\&UFS configuration that satisfies the current latency budget.
This \textit{``just-right''} tuning minimizes per-function power consumption without violating assigned latency targets, enabling SLO-aware and power-efficient execution.

\makeatletter
\newcommand*{\inlineequation}[2][]{%
  \begingroup
    \refstepcounter{equation}%
    \ifx\\#1\\%
    \else
      \label{#1}%
    \fi
    \relpenalty=10000 %
    \binoppenalty=10000 %
    \ensuremath{%
      #2%
    }%
    ~\@eqnnum
  \endgroup
}

%% file: sections/04a_okypous_design.tex
\section{$\boldsymbol{\Omega}$kypous Design}
\label{sec:design}
To realize the above, and based on the opportunities and challenges identified in \S\ref{sec:challenges}, we design $\Omega$kypous: an SLO-aware power management framework for serverless workflows.
Figure~\ref{fig:okypous-design} shows an overview of $\Omega$kypous design.
Overall, $\Omega$kypous comprises four key components.
\textbf{First}, $\Omega$kypous employs a latency budgeting mechanism (\S\ref{subsec:latency-budgeting}) that operates in two phases: an initial phase that statically assigns nominal budgets based on the workflow's structure and target SLO, and a runtime phase implemented by an adaptive controller that dynamically adjusts these budgets based on timing slacks.
\textbf{Second}, it employs a global grey-box performance prediction model (\S\ref{subsec:latency_and_power_prediction}) that estimates function latency across different inputs and CFS\&UFS configurations.
\textbf{Third}, the controller (\S\ref{sec:controller}) uses these latency predictions to select the lowest-power CFS-UFS configuration that satisfies the current function’s budget.
\textbf{Last}, a conflict resolver (\S\ref{sec:conflict-resolver}) mitigates frequency drifts caused by functions sharing the same Core and Uncore components.

\subsection{Latency Budgeting}
\label{subsec:latency-budgeting}
\subsubsection{Pre-deployment Latency Budgeting}
\label{subsubsec:predeployment-latency-budgeting}
\begin{figure}[t]
    \centering
    \includegraphics[width=0.98\columnwidth]{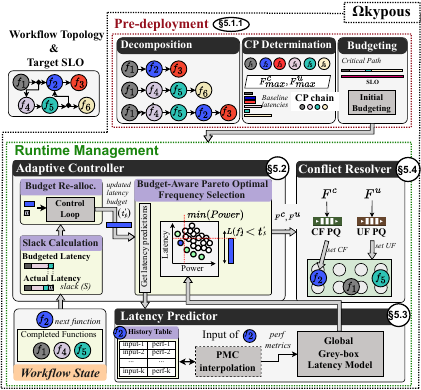}
     \caption{$\Omega$kypous overview.}
     \label{fig:okypous-design}
\end{figure}

To project the end-to-end SLO constraint $T$ onto individual functions, $\Omega$kypous performs a pre-deployment latency budgeting step that assigns nominal latency targets $t_{f_k}$ to each function $f_k$ in the workflow.
Since serverless workflows often contain dynamic control flow (e.g., data-dependent branches), $\Omega$kypous first decomposes the workflow graph into all possible mutually exclusive paths, as shown in Fig.~\ref{fig:okypous-design} (left part of the pre-deployment phase).
For each path $i$, it estimates the baseline latency $l^i_{f_k}(F^c_{\max}, F^u_{\max})$ of every function $f_k$ by executing it in isolation at the maximum CFS-UFS configuration. 
These baseline values reflect the best-case performance achievable by each function when executed in isolation.
To ensure SLO-compliance under worst-case execution scenario, $\Omega$kypous constructs a critical path (CP) estimate as:
\begin{equation}
    CP = \max_i{\sum{l_{f_k}^i(F^c_{max}, F^u_{max})}}
\end{equation}
This represents the path with the highest cumulative baseline latency across all mutually exclusive chains in the workflow.
Notably, the critical path is not necessarily the structurally longest chain (i.e., in number of functions), but rather the one with the highest execution cost based on minimum-latency estimates.
If the SLO constraint $T$ exceeds the critical path estimate ($T > CP$), the additional slack is redistributed across functions using $\Omega$kypous’ latency model (\S\ref{sec:grey-box}), assigning optimized CFS-UFS settings that minimize power while satisfying the end-to-end constraint.
Functions on non-critical paths receive looser latency targets and can operate at lower frequencies, improving energy efficiency. 
Last, for functions shared between the CP and other paths, $\Omega$kypous conservatively selects the tighter latency target, ensuring consistency in SLO adherence across all execution scenarios.

\subsubsection{Dynamic Latency Budgeting}
\label{subsubsec:dynamic-latency-budgeting}
At runtime, $\Omega$kypous updates the latency budget of each function immediately before execution.
The update logic is based on the accumulated timing slack up to the current point.
This slack is not redistributed globally across the remaining workflow but is instead used to update the latency budget of the next function in the execution path.
Specifically, as each function completes, the system computes the residual latency budget by subtracting the cumulative execution time from the already-defined budget so far.
If previous functions complete faster than expected (positive slack), the resulting positive slack is propagated forward, enabling lower-frequency execution for downstream stages.
Conversely, when earlier stages run slower than expected (negative slack), the system tightens the remaining latency budgets to preserve the SLO.
Unlike speculative approaches that forecast entire execution paths~\cite{stojkovic2023specfaas}, $\Omega$kypous adopts a conservative approach that incrementally adjusts latency targets to satisfy the accumulated latency budget up to the current stage.
This adjustment is handled by an adaptive strategy implemented within $\Omega$kypous' controller.

\subsection{Adaptive Controller}
\label{sec:controller}
The controller's objective is to track and adaptively respond to timing slacks propagated from preceding functions; leveraging positive slack to save power and mitigating negative slack to avoid SLO violations.
It follows a closed-loop design, activating control decisions between function invocations.
At each stage, the controller measures the total accumulated slack within the workflow and updates the latency budget for the next function accordingly.
It then queries the latency prediction model (\S\ref{subsec:latency_and_power_prediction}) to select the CFS-UFS configuration that minimizes power consumption while satisfying the adjusted latency constraint.

\noindent$\blacktriangleright$ \textbf{Slack calculation:}
Before invoking each function $f_k$, the controller computes the accumulated latency of all upstream functions: 
\begin{equation}
    L(k{-}1) = \sum_{i=1}^{k{-}1} l_{f_i}
\end{equation}
where \(l_{f_i}\) denotes the actual execution time function \(f_i\). 
Then, it calculates the accumulated slack as $S(k-1) = T(k-1) - L(k-1)$
where the cumulative target latency budget is $T(k{-}1) = \sum_{i=1}^{k{-}1} t_{f_i}$ and \(t_{f_i}\) is the pre-deployment latency allocation for function \(f_i\) (\S\ref{subsubsec:predeployment-latency-budgeting}).
The slack $S(k-1)$ is passed to the adaptive controller, which adjusts and re-allocates the latency budget of the next function $f_k$ based on the proximity to an SLO violation.

\noindent$\blacktriangleright$ \textbf{Adaptive budget re-allocation:}
$\Omega$kypous dynamically adjusts the latency budget of each function at runtime using a feedback controller.
In the context of our control loop, $L$ forms the process variable, $T$ is the setpoint, and $S$ forms the error.
We design our controller according to two principles: \textit{i)} strictly prevent cumulative SLO violations ($S(k) \le 0$), and ii) minimize the total unexploited latency slack over time, formalized as ($\min{\sum_{i=1}^{k}{S(i) \times l(i)}}$).
To achieve this, we maintain near-zero slack between consecutive function stages.
The controller output is defined as $u(k)=K_{p}(S)\times S(k-1)$, where $K_p(S)$ is a dynamically adjusted proportional gain governed by a piecewise-defined function.
For negative slack values (i.e., when latency overshoots the target), $K_p$ is set to a large constant to aggressively recover the lost budget.
For non-negative slack, $K_p$ varies with the magnitude of $S(k-1)$:
when slack is small (close to zero), the controller is conservative, limiting slack propagation to account for prediction errors or runtime noise.
As slack grows, the controller becomes more permissive, allowing greater exploitation to reduce power consumption.
The controller's output $u_{f_k}$ is then used to update the latency budget of the next function $f_k$: $t'_{f_k} = u({f_k}) + t_{f_k}$, where $t_{f_k}$ is the latency budget of function $f_k$ as set in the pre-deployment phase (\S\ref{subsubsec:predeployment-latency-budgeting}).

\noindent$\blacktriangleright$ \textbf{CFS-UFS selection:}
Finally, the controller maps the updated latency budget to an appropriate CFS-UFS configuration.
To do so, it uses a latency prediction model (\S\ref{subsec:latency_and_power_prediction}) to estimate the execution time of previously unseen frequency configurations (if any).
For power estimation, since system power is predominantly influenced by Core and Uncore frequency levels (Fig.~\ref{fig:cfs-ufs-latency-power-analysis}), we use a lightweight second-degree polynomial regression model, which achieves a MAPE of 1.9\%.
From the model-predicted latency-power Pareto frontier, the controller selects the most power-efficient CFS-UFS configuration that satisfies the updated latency budget $t'_{f_k}$, solving the following optimization problem:
$\min{\hat{P}(F^c, F^u)} \text{ s.t. }  \hat{L}(F^c, F^u) \leq t'_{f_k}$.
The final CFS-UFS configuration is given to the Conflict Resolver (\S\ref{sec:conflict-resolver}).

%% file: sections/04b_okypous_modeling.tex
\subsection{Latency Predictor}
\label{subsec:latency_and_power_prediction}
This component estimates function latency under varying Core and Uncore frequency configurations.
To support few-shot scenarios and diverse function behavior evident in serverless environments (\S\ref{sec:challenges}), $\Omega$kypous adopts a global modeling strategy that generalizes across functions and input types.
This eliminates the need for per-function training, but introduces new challenges:
\begin{enumerate}[nosep,leftmargin=*]
    \item \textbf{Unified input representation:} Inputs vary across functions—in type, structure, and scale. Mapping them to a shared representation is non-trivial~\cite{azure2023autonomousdcsmain,masouros2024sparkle}. For example, categorizing payloads into coarse-grained buckets (e.g., labeling datasets under 1MB as “small”) results in a discrete feature space that lacks the flexibility to interpolate to unseen inputs, reducing generalization.
    \item \textbf{Generalizable \& Low-overhead inference:} While global black-box methods (e.g., DNNs~\cite{masouros2024sparkle}, Bayesian Optimization~\cite{zhou2022aquatope}) offer expressive modeling capabilities, they often require extensive training data and incur high inference overheads. In serverless systems—where predictions must occur in the critical path—such latency is prohibitive. This calls for lightweight models that balance generalization with runtime efficiency.
\end{enumerate}
To this end, $\Omega$kypous adopts a lightweight two-stage grey-box modeling approach based on hardware performance counters (PMCs)\cite{wang2022characterizing,bianchini2020toward,masouros2023adrias,masouros2020rusty}, combining first-principles insights with regression-based learning.
This hybrid design improves generalization while maintaining interpretability and low inference cost.

\subsubsection{Unified Input Representation} 
\label{subsubsec:interpolation}
\begin{figure}[t]
	\centering
    \centering
    \includegraphics[width=\columnwidth]{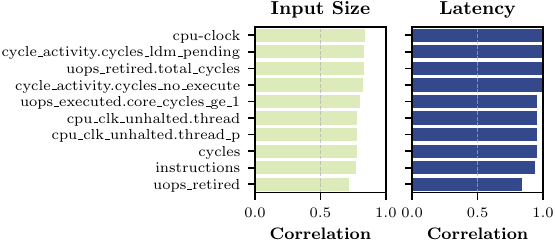}
     \caption{Pearson Correlation of Performance Monitoring Counters (PMCs) with input size and execution time.}
     \label{fig:pmc-correlation}
\end{figure}

To enable a uniform representation for global modeling, $\Omega$kypous maps all functions and their varying inputs to hardware PMCs, collected under the maximum CFS-UFS configuration $(F^c_{\max}, F^u_{\max})$.
Here, input size refers to any quantifiable request feature (e.g., payload size, number of graph nodes) that impacts execution time. 
This mapping provides a function-agnostic abstraction of input characteristics, as PMCs correlate strongly with both input size and execution time.
Figure~\ref{fig:pmc-correlation} shows the average Pearson correlation~\cite{cohen2009pearson} between the top ten PMCs and input size/execution time across our examined functions (\S\ref{sec:experimental-setup}). 
Despite differences in functionality and runtime behavior between various functions, several PMCs exhibit high correlation coefficients ($>0.7$) with input size, indicating their effectiveness as proxies for input-driven latency variation.
Notably, the most correlated events (e.g., \texttt{\small cpu-clock}, \texttt{\small cycles}, \texttt{\small instructions}) reflect core-level activity that scales with workload size and captures the computational volume induced by different inputs.
To generalize to unseen inputs, $\Omega$kypous maintains a \textit{History Table} that logs PMC values from past executions.
When a new input is encountered, the system estimates its PMCs via linear interpolation over nearby entries in the table.
This achieves high accuracy with a MAPE of $1\%${-}$5\%$.

\noindent$\blacktriangleright$ \textbf{Handling Ambiguous Inputs:}
While PMC interpolation based on input size is effective in most cases, inputs of identical size can still exhibit different execution behavior—e.g., when embedded parameters or control fields (such as in JSON) affect runtime.
In such cases, relying solely on input size to estimate PMCs can be misleading. 
To address this, $\Omega$kypous maintains separate history tables for each distinct configuration, identified using available request metadata such as hyperparameters or user-specified flags. 
For cases where the hyperparameters are intractable, automatic feature selection methods can be utilized~\cite{retail}.

\subsubsection{Global grey-box latency modeling}
\label{sec:grey-box}
After estimating the PMCs for the invoked function, we use them to predict the execution latency under different CFS-UFS configurations.
$\Omega$kypous adopts a grey-box modeling approach ~\cite{bohlin2006practical,kroll2000grey} that combines \textit{first-principles system modeling} and \textit{data-driven regression}.
Building upon prior work~\cite{choi2004fine,deng2012coscale,sarood2014maximizing}, we model latency as the sum of cycle counts in the Core ($W^c$), Uncore ($W^u$), and DRAM ($W^d$) domains, each divided by its corresponding frequency ($F^c$, $F^u$, $F^d$):
\begin{equation}
    L(F^c,F^u)=\frac{W^{c}}{F^c} + \frac{W^{u}}{F^u} + \frac{W^{d}}{F^d}
    \label{eq:cycles-to-freq}
\end{equation}

\begin{figure*}[t]
  \centering
  \begin{minipage}[t]{0.4\textwidth}
    \centering
    \includegraphics[width=\linewidth]{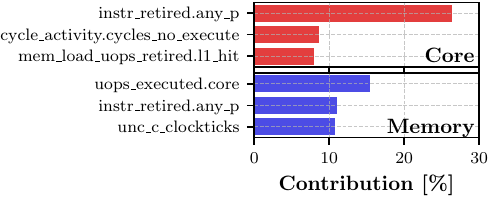}
    \caption{PMCs' relative contribution(\%) for Core, and Memory (Uncore + DRAM) terms in Eq. \ref{eq:linear-model}.}
    \label{fig:pmc-contributions}
  \end{minipage}
  \hfill
  \begin{minipage}[t]{0.34\textwidth}
    \centering
    \includegraphics[width=\linewidth]{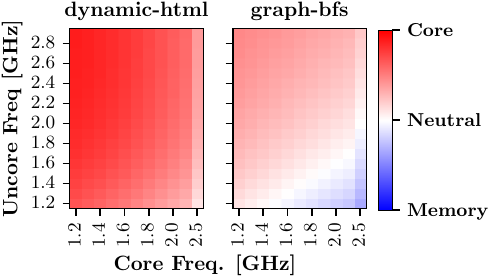}
    \caption{Core/Memory terms contribution to overall prediction for varying CFS, UFS.}
    \label{fig:core-uncore-contribution}
  \end{minipage}
  \hfill
  \begin{minipage}[t]{0.24\textwidth}
    \centering
    \includegraphics[width=\linewidth]{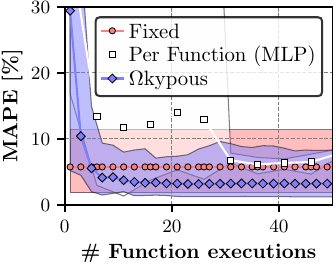}
    \caption{MAPE over Function Invocations.}
    \label{fig:mape-over-time}
  \end{minipage}
\end{figure*}

\noindent Since directly measuring the number of cycles spent in the Core, Uncore, and DRAM domains is not feasible on modern superscalar processors, we approximate each term as a linear combination of hardware PMCs\footnote{Due to hardware limitations (\S\ref{sec:experimental-setup}), DRAM frequency is fixed; its contribution is treated as constant.}.
Specifically, we associate each architectural domain with a set of PMCs and express the corresponding cycle count as a weighted sum of those counters.
Substituting these approximations into Eq.~\ref{eq:cycles-to-freq} yields:
\begin{equation}
L(F^c,F^u)=
\frac{\sum_{i=1}^{N}{c_i\cdot PMC^c_i}}{F^c} + \frac{\sum_{i=1}^{N}{u_i\cdot PMC^u_i}}{F^u} + \sum_{i=1}^{N}{d_i\cdot PMC^d_i}
\label{eq:linear-model}
\end{equation}
Here, $PMC^x_i$ denotes a performance counter $i$ from domain $x \in {c, u, d}$, and $c_i$, $u_i$, $d_i$ are the coefficients associated with each counter.

\noindent$\blacktriangleright$ \textbf{Model training:}
To estimate $c_i$, $u_i$, and $d_i$ in Eq.\ref{eq:linear-model}, we train a single, function-agnostic model using observations from multiple function executions at varying CFS\&UFS.
Each training sample consists of a PMC vector collected under the maximum frequency, the actual CFS-UFS configuration used during execution and the corresponding observed latency. 
The regression fits the analytical model in Eq.\ref{eq:linear-model}, where frequency varies across samples while PMC features remain fixed. 
This allows the model to distinguish the frequency-dependent latency effects from cycle-level behavior encoded in the PMCs.
Frequency is not treated as an input feature; rather, it is applied analytically as a scaling factor during both training and inference.
After evaluating multiple regression techniques, we adopt linear regression due to its lower MAPE ($\approx4\%$) across different functions and inputs.

\noindent$\blacktriangleright$ \textbf{Model explainability:} To interpret the internal behavior of $\Omega$kypous's latency predictor, we analyze the structure of the grey-box model and its learned PMC contributions.
Figure~\ref{fig:pmc-contributions} shows the top PMC features contributing to the Core and Memory (Uncore+DRAM) terms in Eq.\ref{eq:linear-model}.
\texttt{\small instr\_retired.any\_p} dominates Core-side predictions, reflecting total retired instructions, which is a direct indicator of compute activity.
On the memory side, the largest contributor is \texttt{\small uops\_executed.core}, suggesting that memory-induced stalls manifest in reduced execution throughput at the core level.
While Uncore-specific counters like \texttt{\small unc\_c\_clockticks} contribute less, they still provide complementary signals about memory subsystem activity.
These results contrast with the PMCs most correlated with input size and latency (Fig.~\ref{fig:pmc-correlation}), which are dominated by general activity counters—highlighting that effective latency prediction relies on domain-specific signals. We also analyze the relative contributions of Core and Uncore+DRAM latency terms of Eq.~\ref{eq:linear-model} across the CFS–UFS space.
Figure~\ref{fig:core-uncore-contribution} shows results for two functions, \texttt{\small dynamic-html} and \texttt{\small graph-bfs}, where red denotes Core-dominated latency and blue indicates Uncore+DRAM-dominated latency.
For \texttt{\small dynamic-html}, the Core term consistently accounts for over 80\% of predicted latency, even at low UF, indicating compute-bound behavior and potential for Uncore power savings.
In contrast, \texttt{\small graph-bfs} is memory-bound at low UF (up to 70\% from Uncore+DRAM), but shifts to Core-dominated latency as UF increases—demonstrating the model’s sensitivity to changing architectural bottlenecks.

\subsubsection{Runtime Latency Estimation:} At runtime, if no matching input is found in the History Table, $\Omega$kypous estimates the PMC vector via interpolation (\S\ref{subsubsec:interpolation}) and uses it in Eq.~\ref{eq:linear-model} with the target CFS–UFS to compute latency.
The inference requires only a weighted sum and frequency scaling,  incuring only $\approx$3ms to evaluate all CFS-UFS configurations.
As new frequency-latency pairs are observed, $\Omega$kypous incrementally updates the coefficients of Eq.~\ref{eq:linear-model} using the fixed PMC vector, thus, improving the model's accuracy.

\noindent$\blacktriangleright$ {\textbf{Runtime model evaluation:} We compare against two alternatives: \textit{i)} a per-function MLP model inspired by~\cite{stojkovic2024ecofaas}, and \textit{ii)} a static interpolation baseline similar to~\cite{gemini2020zhou}.
Using a function-level leave-one-out setup, the model is incrementally updated as new invocations of the held-out function arrive.
Figure~\ref{fig:mape-over-time} shows that $\Omega$kypous consistently achieves lower error, converging below 4\% MAPE within 5 invocations. 
The MLP, per-function approach performs poorly under data scarcity, with high initial error that reaches a plateau of $\approx$6\%x MAPE after $\approx$30 invocations.
Fixed baseline shows stable but suboptimal performance ($\approx$6\% MAPE on average), and does not benefit from additional samples.
These results highlight the few-shot adaptability of $\Omega$kypous global grey-box model and its ability to generalize across functions.

%% file: sections/04c_okypous_resolver.tex
\subsection{Conflict Resolver}
\label{sec:conflict-resolver}

The Conflict Resolver component is designed to manage DVFS conflicts that arise when co-located functions have differing frequency requirements.
Conflicts most commonly occur within the shared Uncore components on a per-socket basis but can also emerge in scenarios where different functions are co-located on the same physical cores.
One approach to mitigate such conflicts is to co-locate functions with similar frequency needs within the same core pools, as explored in prior work~\cite{stojkovic2024ecofaas}. 
In contrast, $\Omega$kypous adopts a more fine-grained approach by enforcing per-function CFS-UFS settings. 
To handle conflicts, the Resolver uses a priority-based frequency scaling mechanism, maintaining separate priority queues for CFS and UFS requests.
In these queues, priorities are assigned based on frequency values, with higher frequencies receiving higher priority.
When a new function arrives, the Adaptive Controller (\S \ref{sec:controller}) submits a new CFS-UFS configuration to the priority queue.
If the frequency of the submitted request is lower than the currently applied one, the Resolver will retain the current operating frequency.
Otherwise, the newly requested frequency will be applied.
Notably, the Conflict Resolver can effectively manage the co-location of functions with heterogeneous vCPU quota allocations on shared DVFS components.

\subsection{Discussion}
\label{sec:discussion}
$\diamond$ \textit{How does $\Omega$kypous handle uncertainty due to mispredictions, system noise, and queuing effects?}
Besides latency mispredictions, resource interference from application co-location can cause performance variability, adding latency overhead.
This overhead can impact the overall end-to-end latency of the serverless workflow. 
$\Omega$kypous' adaptive controller reactively mitigates this latency overhead, incorporating it into slack calculations.
While this work does not focus on modeling interference, $\Omega$kypous could be used alongside with retraining strategies~\cite{retail} when data drift is detected.

\noindent$\diamond$ \textit{How does $\Omega$kypous handle cold start effects?}
The Azure Trace reveals that the execution time of functions on their premises varies from seconds to minutes~\cite{zhang2021faster,shahrad2020serverless}.
In $\Omega$kypous, we proactively invoke non-existing containers when the first function in the chain is triggered.
This strategy, combined with the adaptive controller (\S \ref{sec:controller}) preserves SLOs.
For workflows with short-running functions (10$s$ of $ms$), $\Omega$kypous can also adopt cold start latency mitigation techniques, e.g., predictive container pre-warming \cite{zhou2022aquatope, roy2022icebreaker, fuerst2021faascache, cai2024incendio}.

\noindent$\diamond$ \textit{Does $\Omega$kypous strategy conflict with the current serverless billing model?}
Ideally, $\Omega$kypous would best fit with an SLO-driven policy that bills clients based on the strictness of the selected SLO requirements.
Yet, modern CSPs usually bill clients proportionally to the time elapsed and the request count~\cite{azure-durable,awslambda}.
Additionally, several serverless workflow offerings are charging based on stage transition count~\cite{awsstep,gcpwf,alibaba-swf}, with some solutions even starting to consider core frequency as a billing factor~\cite{google-freq}.
$\Omega$kypous delivers reduced frequency settings (as shown in \S\ref{sec:charact}) that would be directly translated to reduced costs aligned to billing policies~\cite{google-freq}.
In any case, we note that $\Omega$kypous achieves further improvements in power efficiency under iso-latency scenarios, as shown in Figure \ref{fig:isolatency}, thus reducing data centers' operational expenses.

%% file: sections/04d_okypous_implementation.tex
\subsection{Implementation}
\label{subsec:okypous-implementation}
We implement $\Omega$kypous as an extension of Apache OpenWhisk~\cite{openwhisk-site} over Kubernetes~\cite{kubernetes}. 
Function inputs and outputs are stored in a MinIO~\cite{minio} object store deployed on a remote node to account for realistic network I/O overheads.
To control DVFS, we deploy agents on each physical server that manage Core and Uncore frequency settings.
Core DVFS is configured using the \textit{acpi-cpufreq} driver in \texttt{userspace} mode; system call latencies for frequency changes range from $30-70\mu$s, in line with the $40\mu$s reported in~\cite{gemini2020zhou}.
To ensure frequency-to-function isolation, each VM's vCPU is pinned to a dedicated physical core, and functions are mapped to cores using \texttt{taskset}. 
Uncore frequencies are adjusted through model-specific register \texttt{\small MSR\_UNCORE\_RATIO\_LIMIT} (in address \texttt{\small 0x620}), with a switching latency of 1–3ms.
Power consumption is sampled every 20ms using Intel’s RAPL interface~\cite{pcmgit,david2010rapl}. 
For gathering PMCs, we deploy monitoring agents on each server, which are externally triggered and use \texttt{linux-perf} to capture PMCs across all cores.

%% file: sections/05a_evaluation_setup.tex
\section{Evaluation}
\label{sec:eval}



\subsection{Experimental Setup}
\label{sec:experimental-setup}


\noindent$\blacktriangleright$ \textbf{Examined Infrastructure:} We deploy four VMs with 10 vCPUs and 16GB each, on top of two Intel\textsuperscript{\textregistered} Xeon\textsuperscript{\textregistered} E5-2658A servers, a common setup found in public serverless solutions~\cite{wang2018peeking}.
We pin each VM on a separate socket to allow for granular Uncore frequency, as described in \S\ref{subsec:okypous-implementation}.

\noindent$\blacktriangleright$ \textbf{Examined Serverless Workflows:}We evaluate nine functions drawn from the Serverless Benchmark Suite~\cite{sebs}. Each function is configured with an upper limit of 2 vCPUs, consistent with the median deployment profile reported in~\cite{joosen2023does}. 
This limit does not enforce exclusive core allocation, thereby allowing scenarios involving core co-location (see Section~\ref{sec:conflict-resolver}). 
To simulate realistic workflow compositions, we generate end-to-end workflows by permuting these functions into chains of varying lengths, following structural statistics from~\cite{zhang2021faster}. The resulting workflows include a diverse set of topologies, such as sequential chains, fan-out/fan-in graphs, and conditionally branched workflows.

\noindent$\blacktriangleright$ \textbf{Workflow SLO determination:}
To determine the target SLO for each workflow, we first profile all individual functions under Linux’s \textit{ondemand} frequency governor.
Similar to ~\cite{chen2019parties}, we incrementally increase the request rate (RPS) until reaching saturation, which is defined as the point where 95th percentile (p95) tail latencies begin to rise sharply.
The SLO target for a workflow is then defined as the sum of the per-function p95 tail latencies. Similarly, the maximum supported load for a workflow is conservatively determined by the function with the lowest saturated RPS. 

\noindent$\blacktriangleright$ \textbf{Load generation:} 
Workload is generated using invocation traces sourced from the Azure Public Dataset \cite{zhang2021faster}. 
A workflow invocation is defined as the initial invocation of the first function in the corresponding workflow chain.
To ensure that the load is feasible under our profiled configurations, we downscale the trace’s invocation rate to remain within the maximum RPS thresholds obtained during saturation profiling.
Last, to evaluate performance across representative time intervals in the trace, we applied time series K-means clustering and selected 20-minute invocation windows from samples spanning ten clusters.

%% file: sections/05b_evaluation_results.tex
\subsection{$\Omega$kypous Against Existing DVFS Approaches}
\label{sec:eval-dvfs}

We compare $\Omega$kypous against six baselines, including:
\begin{enumerate}[nosep,leftmargin=*]
    \item \textbf{Linux built-in governors:} We evaluate two Linux governors; \textit{i) performance}, which always selects the highest available frequency, serving as a power-unaware upper bound and \textit{ii) ondemand}, which adjusts frequency based on CPU utilization.
    \item \textbf{DVFS schemes for serverless workflows:} We also consider two prior DVFS mechanisms tailored for serverless workflows; \textit{i) DVFaaS}~\cite{tzenetopoulos2023dvfaas}, which uses a reactive PID controller to regulate Core frequency in response to latency deviations; and \textit{ii) EcoFaaS}~\cite{stojkovic2024ecofaas}, which assigns function invocations to frequency-isolated core pools based on latency constraints. 
    We implement \textit{EcoFaaS} using frequency increments of 400MHz and up to four pools per VM. 
    Pool sizes and frequencies are updated in an interval of $5s$, as mentioned in the paper.
    
    \item \textbf{DVFS schemes for microservices:} Lastly, we compare $\Omega$kypous against two DVFS frameworks originally designed for microservices: \textit{Gemini}~\cite{gemini2020zhou} and \textit{ReTail}~\cite{retail}. 
    Both Gemini's and ReTail's controllers operate per microservice (function in our case), but they do not consider inter-function timing slacks or UFS.
    Gemini uses a NN model per microservice to predict request latency at a baseline Core frequency (1.4GHz), then extrapolates to other frequencies using a scaling function.
    ReTail uses lightweight per-microservice ML models trained on request-specific features (e.g., payload size, parameter values) to select Core frequency settings.
    To focus the evaluation analysis on the control strategies rather than prediction accuracy, we implement both using oracle-based latency predictors -- representing an upper bound on prediction accuracy.
    For Gemini, we provide the true latency at the baseline frequency and apply its original extrapolation method to estimate latency at other CFS levels. 
\end{enumerate}

\begin{figure}
    \centering
    \subfloat[Latency / SLO target ratio.
    \label{fig:main-results-latency}]{
        \includegraphics[width=.49\columnwidth]{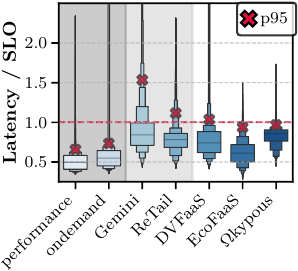}
    }
    \hspace*{\fill}
    \subfloat[Power consumption (W).]{
        \includegraphics[width=0.49\columnwidth]{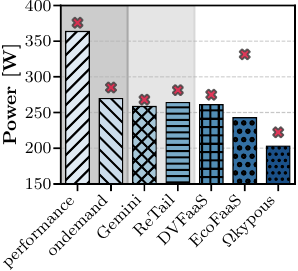}
        \label{fig:main-results-power}
    }
    \caption{Governors' SLO compliance and power efficiency over 160 workflows using the Azure Trace \cite{zhang2021faster}.}
    \label{fig:main-results}
\end{figure}

\begin{figure}
\centering
\begin{minipage}{.4\columnwidth}
  \includegraphics[width=1\columnwidth]{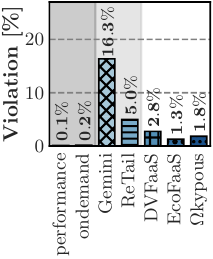}
  \caption{Violation rate.}\label{fig:violations}
\end{minipage}
\hspace{2pt}
\begin{minipage}{.54\columnwidth}
  \includegraphics[width=1\columnwidth]{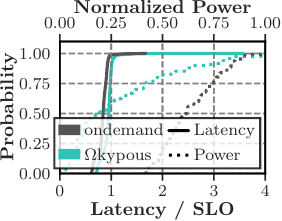}
  \caption{CDF of normalized latency (ratio of observed to the target) - solid line - and normalized power - dashed line.}\label{fig:isolatency}
\end{minipage}%
\end{figure}

\noindent We execute 160 unique workflows with varying lengths, reflecting the distribution observed in the Azure Trace~\cite{zhang2021faster}, invoked according to the trace's workload patterns.
Figure \ref{fig:main-results} demonstrates the latency distribution w.r.t. the SLO target and the average power consumption, while Fig. \ref{fig:violations} shows the SLO violation percentages.

\subsubsection{$\Omega$kypous vs. Linux Governors}
$\Omega$kypous reduces power consumption by 44\% and 25\% compared to \textit{performance} and \textit{ondemand} governors respectively.
\textit{Performance} governor prioritizes execution speed, achieving just 0.1\% SLO violations but at the cost of high power consumption (363W on average).
In contrast, \textit{ondemand} scales frequency based on CPU utilization, reducing power to 269W on average, with only a slight increase in SLO violations.
The demand-driven, but slack-unaware \textit{ondemand} governor often overprovisions under relaxed SLOs (Fig. \ref{fig:main-results-latency}).
To further compare its efficiency with $\Omega$kypous, we repeat the experiment, setting a much stricter SLO target.
Figure~\ref{fig:isolatency} shows the CDF of power (normalized to the maximum) and latency-to-SLO ratio.
By applying CFS-UFS configurations tailored to function profiles, $\Omega$kypous maintains SLO compliance while operating closer to the target (latency/SLO $\approx$1).
With a median latency ratio of approximately 0.95, it achieves a median power reduction of around 30W.
Unlike the \textit{ondemand} governor, $\Omega$kypous leverages predictive modeling to make informed control decisions and exploits UFS, which in some cases achieves iso-latency operation at lower power levels (\S \ref{fig:pareto-analysis}).


\subsubsection{$\Omega$kypous vs. Serverless DVFS schemes}
As shown in Figure~\ref{fig:main-results-power}, $\Omega$kypous achieves the best power–latency balance, reducing power by $\approx$22\% and $\approx$16.4\% compared to \textit{DVFaaS} and \textit{EcoFaaS}, respectively, while maintaining a tighter latency-to-SLO ratio (median $\approx$ 0.86, Fig.~\ref{fig:main-results-latency}).
In terms of SLO violations, \textit{DVFaaS} maintains SLO violations at 2.8\%. 
\textit{DVFaaS} applies reactive PID control without explicit latency modeling. As a result, it resorts to aggressive frequency boosts near the workflow tail to meet SLO targets, causing 2.8\% SLO violations and higher power consumption ($\approx$261W).
\textit{EcoFaaS} selects a Core pool with frequency higher or equal to the desired for each request and reassigns the function accordingly.
While conservative in latency (1.3\% violations), this pool-based approach overprovisions frequency.
Almost half of the functions set their desired frequency to an existing pool,  whereas the rest are executed at elevated frequencies. 
In these cases, the assigned frequencies exceed the desired ones by an average of $\approx$ 375MHz.
Additionally, when high frequencies need to be enforced, the pool-based approach results in increased power peaks (Fig. \ref{fig:main-results-power}).
$\Omega$kypous employing per-function CFS/UFS, avoids the conflicts of a pool-based approach.
Still, a negligible amount of conflicts occur due to shared Core/Uncore components (discussed in \S \ref{sec:charact}).

\subsubsection{$\Omega$kypous vs. Microservice-level DVFS schemes.}
$\Omega$kypous' slack-aware controller achieves up to 15\% fewer violations while achieving $\approx$ 22\% power savings compared to the microservice-level DVFS schemes.
\textit{Gemini} and  \textit{ReTail} manage functions independently, ignoring inter-function slack.
As a result, they neither exploit positive slack nor mitigate negative slack across workflows.
Gemini's reliance on static latency-to-frequency mappings, combined with slack unawareness, leads to 16\% violations.
ReTail improves on this with fewer violations (5\%) but still consumes 258W, as it overlooks the benefits of UFS and positive timing slack exploitation.

\subsection{Power Capping}

\begin{figure}[t]
\centering
\subfloat[Power/Latency distributions.]{
    \includegraphics[width=0.588055\linewidth,keepaspectratio=true]{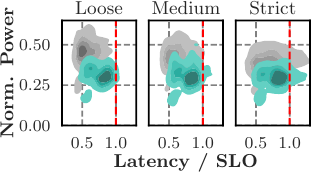}
\label{fig:rapl-kde}}
\subfloat[Violation rate.]{
    \includegraphics[width=0.33\linewidth,keepaspectratio=true]{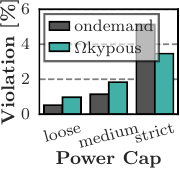}
\label{fig:rapl-violations}}
\caption{Power Capping.}
\label{fig:power-cap}
\end{figure}

\label{sec:capping}
Next, we evaluate the efficacy of $\Omega$kypous under RAPL-based power capping, a technique followed by prior work \cite{lo2014towards} to reduce power.
In Fig. \ref{fig:power-cap}, we illustrate the normalized power, and latency distributions of \textit{ondemand} and $\Omega$kypous, as well as the violation count under three per-socket power capping levels (loose -- 40W, medium -- 42.5W, and strict -- 45W).
We observe that the normalized (min-max) power distribution of
our approach leans towards lower values while being dense in latency values close to the SLO target (Fig. \ref{fig:rapl-kde}).
The violation rate of $\Omega$kypous is consistently lower than 3.5\%.
Notably, the aggressiveness of $\Omega$kypous' controller can be fine-tuned to address different scenarios.
Other governors were unable to meet the SLO, with ReTail resulting in 32\% violations in the looser power cap level.

\subsection{{$\Omega$}kypous characterization}
\label{sec:charact}

\begin{figure}[t]
\centering
\includegraphics[width=0.7\linewidth,keepaspectratio=true]{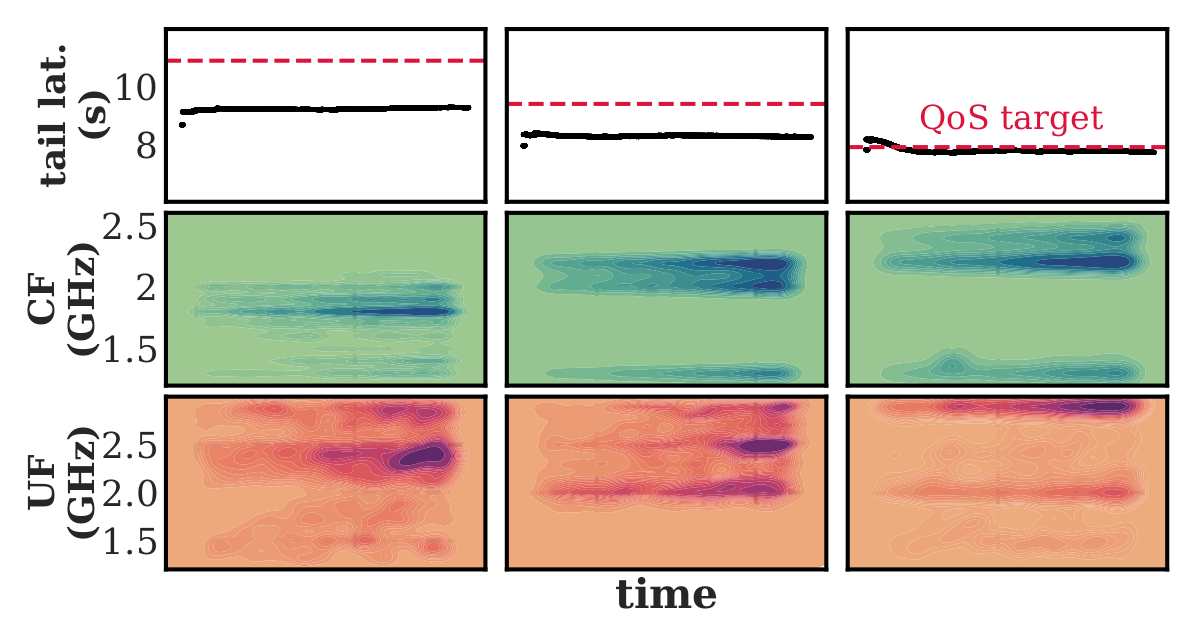}
\caption{Frequency resolution and tail latency.}
\label{fig:bob}
\end{figure}

\noindent$\blacktriangleright$ \textbf{Frequency resolution \& tail latency:}
We explore the operational dynamics of $\Omega$kypous, particularly analyzing its behavior under varying conditions.
We executed a 5-stage workflow with increasing RPS levels (40 to 100\% of the max RPS) and tested against three escalated SLO strictness scenarios. 
Figure~\ref{fig:bob} shows tail latency over time (black line), corresponding to each SLO constraint (indicated by the red lines) along with the fluctuations in Core and Uncore frequencies as RPS increases.
The results demonstrate that tail latency consistently stays below the pre-set SLO thresholds, proving $\Omega$kypous's effectiveness in satisfying diverse scenarios.
CFS and UFS exhibit an upward trend in their distributions both i) as the RPS intensifies and ii) as the SLO target becomes more demanding.
Notably, UFS exhibits higher impacts in latency regulation when satisfying lower SLO constraints.
We repeated this experiment for the \textit{ondemand} governor and observed by 400MHz higher frequency for more than 80\% of the experiment time.


\begin{figure}[t]
\centering
\subfloat[\#cores sharing the same  Core power-frequency  domain.]{
\includegraphics[width=0.49\columnwidth,keepaspectratio=true]{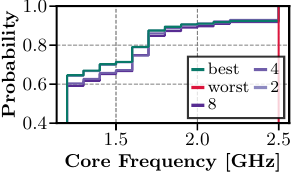}
\label{fig:conflict_core}}
\hspace{0.07cm}
\subfloat[\#cores sharing the same Uncore power-frequency domain.
]{
    \includegraphics[width=0.45\linewidth,keepaspectratio=true]{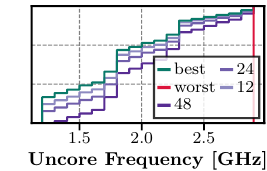}
\label{fig:conflict_uncore}}
\caption{Impact of conflict resolving on frequency drifting.}\label{fig:conflict-impact}
\end{figure}

\noindent$\blacktriangleright$ \textbf{Impact of Conflict Resolution:}
We examine the impact of conflict resolution under scaled Core/Uncore domain sharing scenarios.
Using the requested frequencies from concurrently running functions received by the Conflict Resolver (\S \ref{sec:conflict-resolver}) during the Azure Trace invocations,
we calculate the volume of frequency drifting at the Core -- Fig. \ref{fig:conflict_core}, and Uncore units -- Fig. \ref{fig:conflict_uncore}.
The \textit{best case} (upper limit) assumes each HW thread has a dedicated Core/Uncore domain.
In Fig. \ref{fig:conflict_core} the distributions denoted as 2, 4 and 8 represent scenarios where several HW threads share the same physical Core unit, e.g., due to HyperThreading.
Similarly, Fig. \ref{fig:conflict_uncore} shows the Uncore frequency drifting when 12, 24, or 48 Cores share the same Uncore unit.
The impact of the Resolver on the Core power-frequency domain is negligible (Fig. \ref{fig:conflict_core}).
Specifically, the average difference between the frequency distributions of the best, and the maximum Core sharing (8) is $\approx$ 35MHz.
Concerning the Uncore Unit (Fig. \ref{fig:conflict_uncore}), we observe a higher volume of drifting between the different sharing scenarios.
The average frequency difference between the best and the 48 case, corresponds to $\approx$130MHz.
Yet, this is much more efficient than the unmanaged Uncore (worst case), which selects the maximum frequency.
Notably, emerging Uncore hardware configurations\footnote{For example, Intel Sapphire Rapids high-core count variety (XCC)}.
that promise intra-package Uncore DVFS domains could potentially mitigate the presented frequency drift.




\begin{figure}[t]
\centering
\includegraphics[width=0.75\linewidth,keepaspectratio=true]{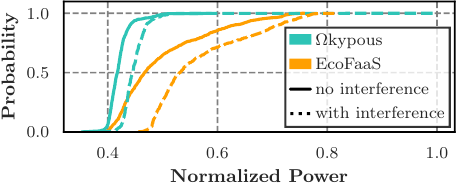}
\caption{Normalized power consumption of serverless workflows executed under interference.}
\label{fig:noise-power}
\end{figure}
\noindent$\blacktriangleright$ \textbf{Controller's Resilience:}
\AT{}
We evaluate the resilience of the $\Omega$kypous controller in the presence of resource interference caused by co-location and suboptimal initial latency budgeting. 
First, to evaluate resilience to interference, we co-locate an XGBoost~\cite{chen2016xgboost} training job on each physical host while concurrently executing twenty workflows using EcoFaaS and $\Omega$kypoys. 
While both frameworks meet their SLOs, they differ in their power consumption behavior.
Figure~\ref{fig:noise-power} presents the CDF of normalized power consumption.
$\Omega$kypous mitigates negative slacks and results in a minimal power increase (214W on average), while EcoFaaS increases to $\approx$ 271W.   
$\Omega$kypous controller also mitigates errors introduced by poor latency budgeting. 
We execute workflows with high latency functions at the beginning (right-skewed) and the end (left-skewed), assigning equal SLO targets per function.
Right-skewed scenarios show 0\% violations at 240W, while left-skewed scenarios result in a 3\% violation rate at 210W.

\noindent$\blacktriangleright$ \textbf{$\Omega$kypous overhead:}
We note that the overheads related to prediction latency, frequency determination, and the time required for the frequency to take effect have all been included in the evaluation experiments.
More in detail, the interpolation and the \textit{perf} metrics pre-processing requires 3-15$ms$, while the inference for latency and power is 2-5$ms$ in total for all the configurations.
The Conflict Resolver requires 10-20$ms$ to apply the desired frequency to the remote node before and/or after the function invocation.
The overhead is reasonable as 75\% of functions in Azure exhibit second-scale latency \cite{shahrad2020serverless}. 
For functions with latency below 100ms (5\%), DVFS can be decided for groups of consecutive functions.
Nevertheless, workflows with both long- and short-running functions can effectively mitigate this overhead through $\Omega$kypous's slack management approach.

%% file: sections/06_conclusion.tex
\section{Conclusion}


We introduced $\Omega$kypous, an SLO-aware DVFS framework for serverless workflows that addresses the growing prominence of latency propagation effects in serverless chains.
$\Omega$kypous employs a grey-box modeling approach to predict execution latency with minimal sampling and guide coordinated CFS and UFS control decisions. 
By jointly adjusting Core and Uncore frequencies, $\Omega$kypous' adaptive controller maintains end-to-end SLO compliance while optimizing for power efficiency.
Evaluated on the Azure Trace \cite{zhang2021faster}, $\Omega$kypous consistently outperforms SotA power managers, achieving $\approx$ 16\% power reduction while minimizing violations.